\documentclass[letterpaper,twocolumn,english,aps,prb,groupedaddress,superscriptaddress,showpacs,amsfonts]{revtex4}
\usepackage{ae}
\usepackage{aecompl}
\usepackage[T1]{fontenc}
\usepackage[latin1]{inputenc}
\usepackage{float}
\usepackage{amsmath}
\usepackage{graphicx}
\usepackage{amssymb}

\makeatletter
\usepackage{babel}
\makeatother
\begin{document}

\title{First-principles Calculations of Engineered Surface Spin Structures}

\author{Chiung-Yuan Lin}

\affiliation{Department of Electronics Engineering, National Chiao
Tung University, Hsinchu, Taiwan}

\author{B.~A.~Jones}

\affiliation{IBM Almaden Research Center, San Jose, CA 95120-6099,
USA}

\date{\today}

\begin{abstract}
The engineered spin structures recently built and measured in
scanning tunneling microscope experiments are calculated using
density functional theory. By determining the precise local
structure around the surface impurities, we find the Mn
atoms can form molecular structures with the binding surface,
behaving like surface molecular magnets. The spin structures are
confirmed to be antiferromagnetic, and the exchange couplings are
calculated within $8\%$ of the experimental values simply by
collinear-spin GGA+U calculations. We can also explain why the
exchange couplings significantly change with different impurity
binding sites from the determined local structure. The bond
polarity is studied by calculating the atomic charges with and
without the Mn adatoms.
In addition, we study a second adatom, Co.
We study the surface Kondo effect of Co by calculating the
surrounding local density of states and the on-site Coulomb $U$,
and compare and contrast the behavior of Co and Mn.
Finally, our calculations confirm that the Mn and Co spins of
these structures are 5/2 and 3/2 respectively,
as also measured indirectly by STM.
\end{abstract}

\maketitle

\section{Introduction}
Assembling and manipulating a few spins (1$\sim$20) is essential for
the development of nano magnetic devices. During the past decades,
chemists have been able to synthesize molecular magnets that
carry giant molecular spins. Potential applications of molecular
magnets have been extensively proposed in the literature \cite{MM},
such as magnetic storage bits, quantum computation, and
magnetooptical switches. The atoms within a molecular magnet form
chemical bonds with each other, and therefore are very difficult
to manipulate. Instead of assembling atomic spins chemically to
form isolated molecules, the advance of manipulating atoms on
surfaces by scanning tunnelling microscope (STM) has made it
possible to make, probe, and manipulate individual atomic spins.

In a pioneering experiment, Hirjibehedin et al. \cite{MnScience}
carried out low-temperature STM measurements of atomic chains of
up to 10 Mn atoms. These magnetic chains are assembled by atomic
manipulation on copper nitride islands that provide an insulating
monolayer between the chains and a Cu(100) substrate (to be called
CuN surface later in this paper).
Ref.~2 shows the calculation of exchange coupling $J$
using the Heisenberg Hamiltonian to be successful.
It demonstrated that the exchange coupling $J$ can be tuned
by placing the magnetic atoms at different binding sites on the
substrate. Nevertheless, the STM experiments can not provide
either a detailed study of the single CuN layer or the sub-atomic
spatial structures around the Mn atoms. As we will show in this
work \cite{Mn2009PRB}, the former can explain why tunnelling current and spin can
both be preserved, and the later is essential for realizing the
molecular magnetism of the Mn-surface complex as well as
understanding how $J$ depends on the Mn binding site. Moreover,
the $5/2$ spin of the Mn atoms on such a CuN surface is calculated
directly here rather than indirectly extracting from
inelastic-tunnelling-spectroscopy steps in the experiments.

In addition to the interatomic magnetic coupling, the surface Kondo effect
is also an interesting topic in engineered spin systems. Recent studies
show that the surface Kondo effect is interestingly influenced by
either the magnetic anisotropy of the Kondo atom itself \cite{CoNaturePhysics}
or by being coupled to a second magnetic atom \cite{CoPRL}.
These systems both have Co as the adatom for their Kondo impurity and are built on the CuN
surface that was previously used to study coupling of Mn atoms.
These experimental studies explain surface Kondo under external influences
using phenological models, and obtain great success. However, detailed
microscopic understandings such as the local density of states (LDOS)
around the Co and the on-site Coulomb repulsion $U$ were not achieved yet.
Also, Ref.~4 concludes indirectly that the Co spin on this surface is
$S=3/2$ by first excluding $S=1/2$ and integer $S$ from the experimental fitting
and then excluding $S\geq5/2$ based on the experience that the spins of
surface-adsorbed atoms are generally unchanged or reduced from the free atom.
Yet a direct measurement or calculation was not done.

In this work, we perform first-principles calculations of the clean
CuN surface and of Mn and Co adatoms on this surface with structure
optimization. We find, surprisingly, that when the Mn atoms are
deposited on the Cu sites of the CuN surface, the nearby N atoms
break bounds with their neighboring Cu and form a ``quasi''
molecular structure on the surface, a situation
which does not happen for Mn
at the N sites. This fact itself was not determined from
experiment, and can only be realized from a first-principles
calculation. As a comparison, we study the clean CuN surface
and find that the CuN monolayer is formed by polar covalently
bounded Cu and N, and such a layer is shown to provide a
semi-metal surface layer on the
underlying Cu substrate allowing the
coexistence of the Mn spin and STM current. We also accurately
calculate the exchange coupling $J$ using the GGA+U method, from
which we demonstrate that first-principles calculation has the
capability of predicting $J$ of given physical systems.
For a Co atom on the same surface, we determined the on-site Coulomb $U$
that is very important in understanding the Kondo effect.
We also compare the LDOS of Co on the Cu and N sites, and
explain why the Kondo effect is observed in the experiments on the
Cu site but not on the N.
Finally, we determine, by analyzing the Co density of states,
a Co spin that matches what was measured indirectly
from STM experiments \cite{CoNaturePhysics}.

\section{Theory}
\begin{figure}
\begin{center}
\includegraphics[keepaspectratio,width=6.5cm]{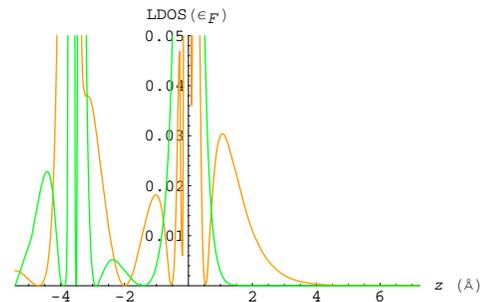}
\end{center}
\caption{\label{fig-CuN-wfn} LDOS($\epsilon_{F}$) along the
out-of-surface direction with the surface Cu atom as the origin,
for both the clean Cu(100) (orange) and CuN (green) surfaces.
(The vacuum corresponds to positive values of $z$.)}
\end{figure}

The CuN monolayer between the magnetic atoms and Cu substrate
originates from the idea of preserving the atomic spins from being
screened by the underlying conduction electrons, while at the same
time allowing enough tunneling current from an STM tip to probe
the spin excitations. To understand this further in a microscopic
picture, we simulate both the Cu(100) and CuN surfaces by a
supercell of 7-layer slabs separated by 8 vacuum layers, where for
the CuN surface, each slab has CuN monolayers on both sides
and three Cu layers in between. The electronic structure is
calculated, in the frame work of density functional theory, using
the all-electron full-potential linearized augmented plane wave
(FLAPW) method \cite{win2k} with the exchange-correlation
potential in the generalized gradient approximation (GGA)
\cite{PBE96}. We calculate the LDOS of
both the Cu(100) and CuN surfaces at the Fermi energy as a
function along the $z$ direction through the surface Cu atom. As
seen from Fig.~\ref{fig-CuN-wfn}, the LDOS of the clean Cu(100)
surface has a much longer tail into vacuum than the CuN surface.
The calculated work functions are 4.6 and 5.2 eV respectively, a
difference of 0.6 eV, much smaller than a typical bulk insulator,
which has a work function $>\!\!\!\!\!\!\!_{_{_{\textstyle \sim}}}\,3$ eV
more than copper. This shows that the CuN monolayer provides the
Cu substrate a moderate conduction that makes possible the coexistence of
the atomic spin and STM current.

\begin{figure}
\begin{center}
\includegraphics[keepaspectratio,trim=4cm 0cm 1cm 3cm,clip,width=8cm]{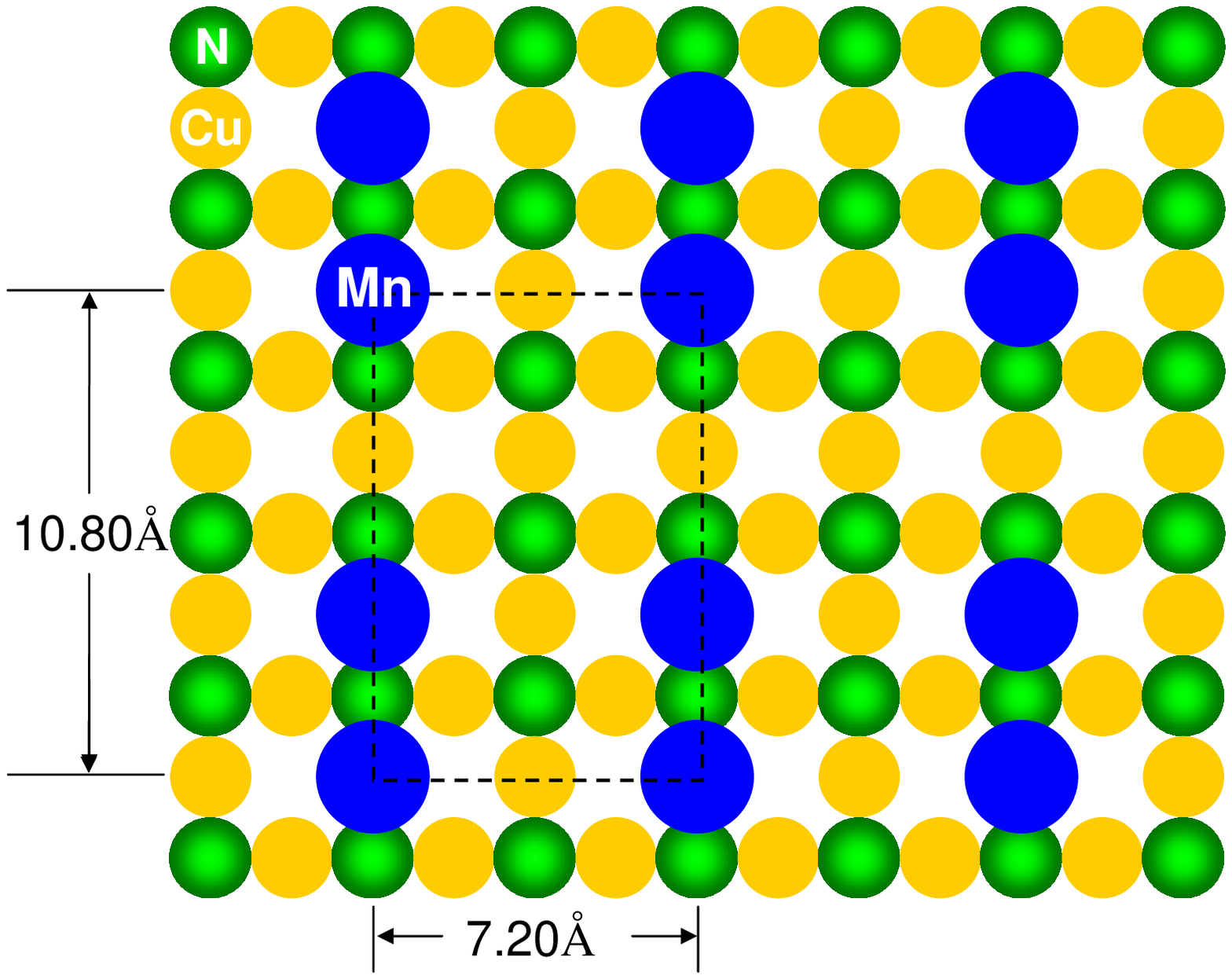}
\end{center}
\caption{\label{Mn2-unitcell} Unit cell of a Mn dimer on the CuN
surface.}
\end{figure}

To calculate the electronic structures of Mn(Co) on the CuN surface,
we simulate the single magnetic atom on this surface by a supercell of 5-layer slabs similar
to the one for CuN surface with the Mn(Co) atoms placed on top of the
CuN surface at $7.24 \AA$ separation. The crystal structure is
optimized until the maximum force among all the atoms reduces to
$<\!\!\!\!\!\!\!_{_{_{\textstyle \sim}}}\,2$ mRy$/a_{0}$. The $3d$
orbital can in general have strong Coulomb repulsion $U$ that can
not be taken into account by GGA. Using a constraint-GGA method
\cite{Novak}, we obtain the $U$ value of a single Mn at the Cu
site of the CuN surface to be 4.9 eV, and 3.9 eV at the N site.
Since the calculated Mn $U$'s fall in the range of strong correlation,
they are then used in the GGA+U calculation
\cite{Anisimov} for Mn $3d$.
To calculate a Mn dimer on the CuN surface,
we simulate the system by the same slab setup as the single Mn
except that the Mn atoms on the surface are arranged as in
Fig.~\ref{Mn2-unitcell}. The electronic structure is also
calculated using GGA+U with $U$ on the Mn $3d$ orbitals.
For Co on the Cu site, we also apply the constraint-GGA method and obtain $U=0.8$ eV.
We then calculate this system by GGA with no additional $U$. In fact,
since the experiments show such a Co adatom exhibits the Kondo effect,
it does not make sense to apply the $U$ statically in a dynamical process (Kondo).

\section{Results and Discussion}

\begin{figure}
\begin{center}
\includegraphics[keepaspectratio,trim=5cm 2cm 1cm 4cm,clip,width=8cm]{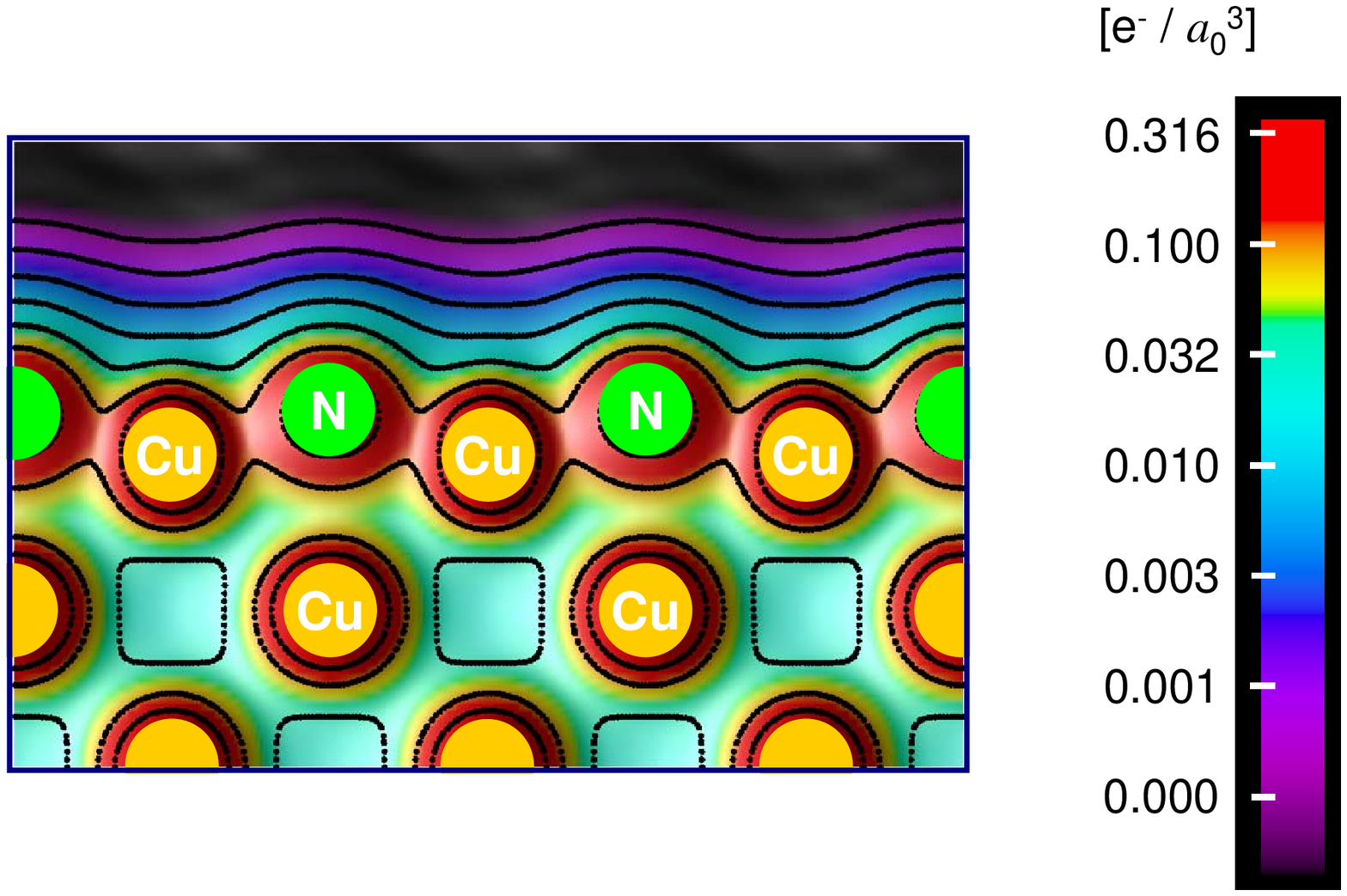}
\end{center}
\caption{\label{fig-CuN-rho} Electron density contour of the CuN
surface along the N raw and the out-of-plane direction.}
\end{figure}

\begin{figure}
\begin{center}
\includegraphics[keepaspectratio,trim=5cm 2cm 1cm 4cm,clip,width=8cm]{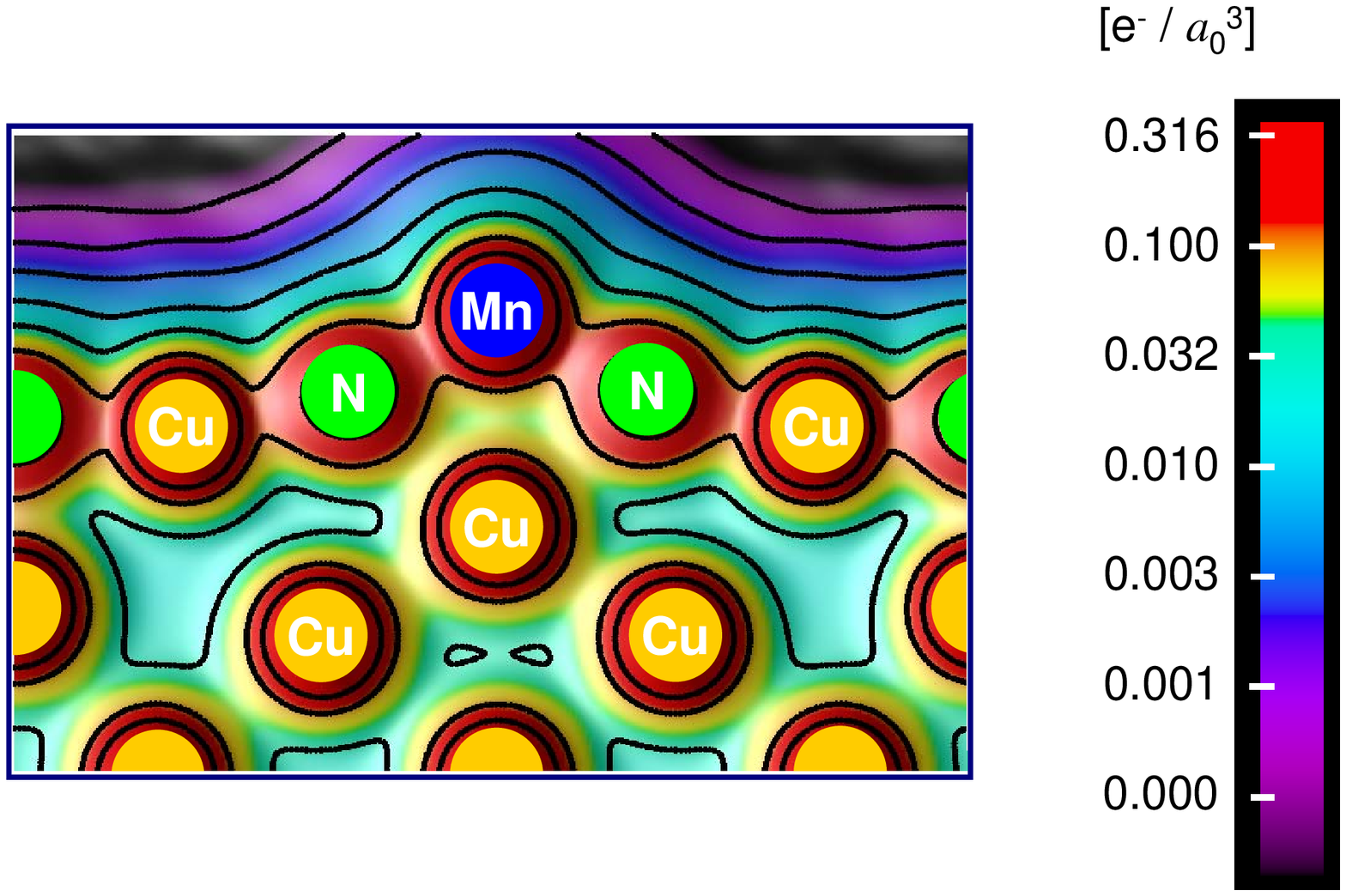}
\end{center}
\caption{\label{fig-MnCuN-rho} Electron density contour of a
single Mn on the CuN surface along the N-Mn-N raw and the
out-of-plane direction.}
\end{figure}

\begin{figure}
\begin{center}
\includegraphics[keepaspectratio,width=6.5cm]{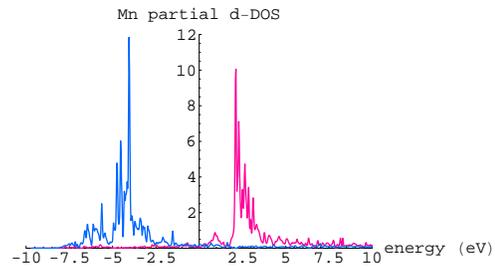}
\end{center}
\caption{\label{fig-MnCuN-dos} Mn $d$-projected density of states
of a single Mn on the CuN surface (the leftmost curve (blue in color)
for spin up and the rightmost curve (pink) for spin
down).}
\end{figure}

\begin{table}
\begin{tabular}{|c|c|c|}
  \hline
   Exchange coupling $J$ & Cu-site Mn dimer & Cu-site Mn dimer \\
                   (meV) &                  &                  \\
  \hline
      GGA$+$U     & $6.50\pm0.05$ & $2.5$       \\
   (calculated $U$)   & ($U=4.9$eV)   & ($U=3.9$eV) \\
  \hline
       STM        & $6.2\pm0.3$   & $2.7$ \\
  \hline
  GGA & $18.5$ & $-1.8$ \\
  \hline
       GGA+U      &     5.4       & 5.1 \\
  (calculated $U$+1eV)&               &     \\
  \hline
\end{tabular}
\caption{\label{J} Calculated exchange coupling $J$ at different
$U$ values, compared with the STM measurements.}
\end{table}

To see the effect on the surface of the presence of Mn atoms, we plot the
electron density of the clean CuN surface in
Fig.~\ref{fig-CuN-rho}. We find the N atoms snug in between the surface Cu
atoms to form a CuN surface layer, joined by shared charge densities as well as proximity.
The vertical distance between N and the
surface Cu is only 0.26 \AA, essentially collinear.
The density contour shared by N and Cu indicates that
a polar covalent bond is formed between Cu (metallic) to N (larger
electronegativity). In fact, a Bader analysis \cite{Bader} on our
calculated electron-density distribution shows N and surface Cu
are $-1.2$ and $+0.6$ charged respectively.
Fig.~\ref{fig-MnCuN-rho} shows the electron density contour of a
single Mn atom placed on the Cu atom on this surface. As one can
see, the atomic structure is substantially rearranged. The Mn atom
attracts its neighboring N atoms remarkably out of the surface,
forming a new polar covalent bond that replaces the CuN binding
network, and the Cu atom underneath Mn moves towards the bulk. We
have calculated that Mn and its neighboring N are $+1.0$ and
$-1.3$ charged respectively, indicating that the Mn-N bond has a
stronger polarity than the Cu-N.

The calculated density of states
for a single Mn on the CuN surface is plotted in
Fig.~\ref{fig-MnCuN-dos}. It is clearly seen that the Mn $3d$
majority spin states are all below the Fermi level and the
minority states are all above, which implies a $3d^{5}$
configuration for Mn, a spin $S=5/2$ configuration. We also do
the same
analysis for Mn at the N site of the CuN surface,
and this structure exhibits the same unchanged Mn spin. This verifies the same
conclusion drawn from comparing spin chains of different
lengths in the STM experiment \cite{MnScience}.

We now consider the exchange coupling of a dimer of Mn.
The spin excitation measured by STM \cite{MnScience}
occurs between the antisymmetric
spin ground state and the first excited state. These quantum
atomic-spin states are not accessible by density-functional
electronic-structure calculation. However, the collinear spin
states (with parallel and antiparallel spins) of a Heisenberg spin
dimer exactly correspond to the collinear magnetic-moment
configurations of the real crystal system of the Mn dimer
absorbed on the CuN substrate. The parallel and antiparallel spin
states have energy expectation values $\pm JS^2$ respectively. One
simply takes the difference of the total energies between the
parallel- and antiparallel-spin dimer on the CuN surface, and
then extract $J$ from this energy difference $\delta E$ and $S$ by
the following equation,
\begin{equation}
\delta E=JS^2-(-JS^2)=2JS^2 \label{EJS}
\end{equation}
For a Mn dimer at the Cu site of a
CuN surface, we obtain an exchange coupling of $J=6.4$ eV from
(\ref{EJS}), which shows excellent agreement with the STM measured
$J=6.2\pm0.2$ eV. In order to show that this agreement is not just
a coincidence, we do the same calculation for a Mn dimer on the
N site. The exchange coupling $J$ turns out to be $2.5$ eV, which
is also close to the STM measurement ($J=2.7$ eV), and is roughly
half of the Cu-site $J$. Thus, we have demonstrated that DFT
reproduces the exchange coupling between these engineered spins,
and will have the capability of predicting similar systems.

In order to check whether it is reasonable to use the $U$ values
determined by the constraint-GGA method in calculating $J$, we
also calculate $J$ using other $U$ values. The resulting $J$'s are
listed in Table \ref{J}. We note the significant lack of the agreement of $J$
calculated by alternative methods with the experimental values.
This strongly suggests that the
constraint-GGA method can very likely be used to correctly predict
the exchange couplings of similar spin systems.

\begin{figure}
\begin{center}
\includegraphics[keepaspectratio,trim=5cm 3cm 1cm 4cm,clip,width=8cm]{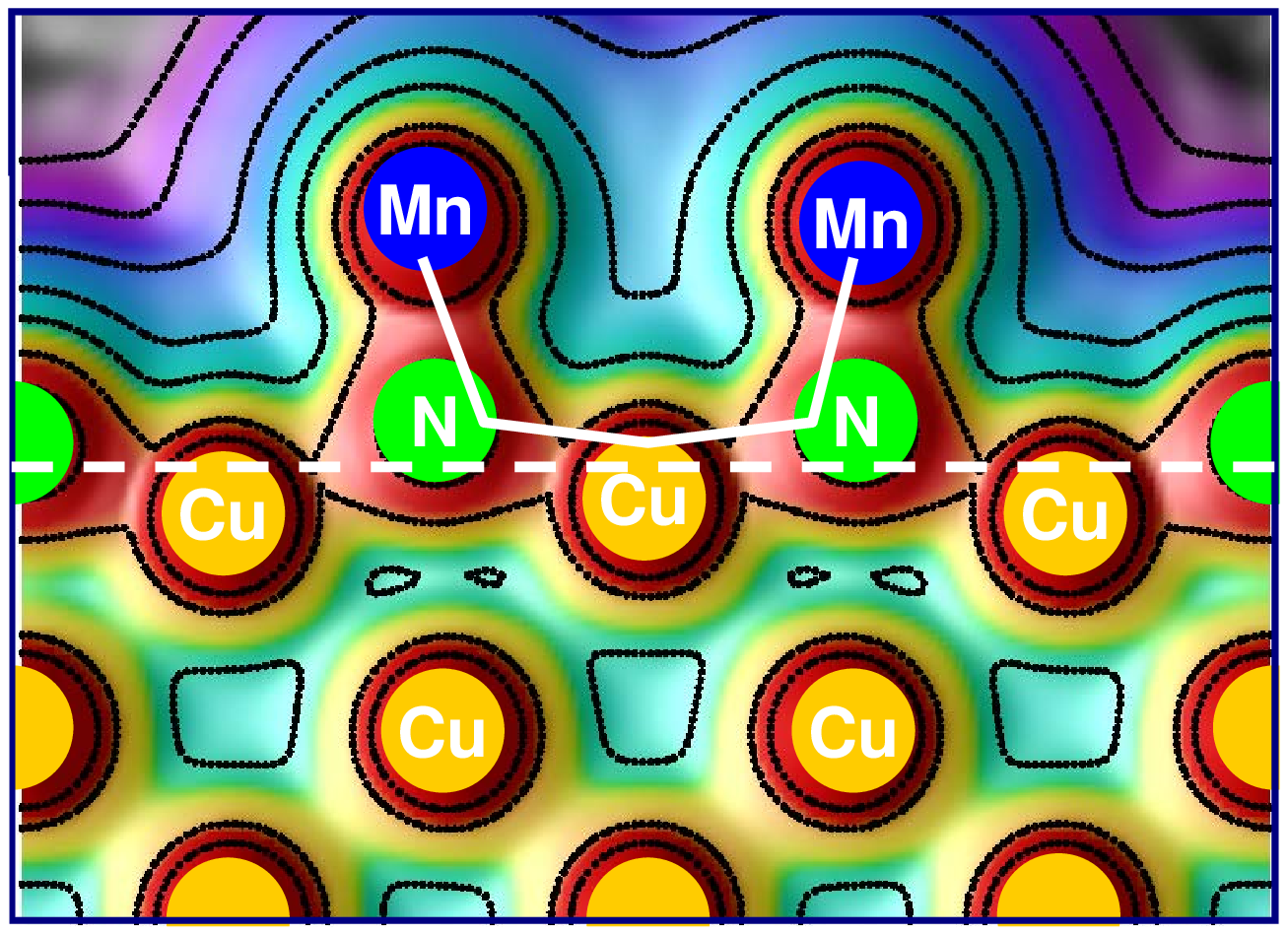}
\end{center}
\caption{\label{fig-Mn2CuN-Nsite-rho} Electron density contour of a Mn
dimer at the N site of the CuN surface along the Mn-N raw and the out-of-plane direction.
The white solid line shows our proposed coupling path between Mn spins.
The dashed shows how corrugated the clean CuN surface becomes in the presence of Mn adatoms.}
\end{figure}

\begin{figure}
\begin{center}
\includegraphics[keepaspectratio,trim=5cm 3cm 1cm 4cm,clip,width=8cm]{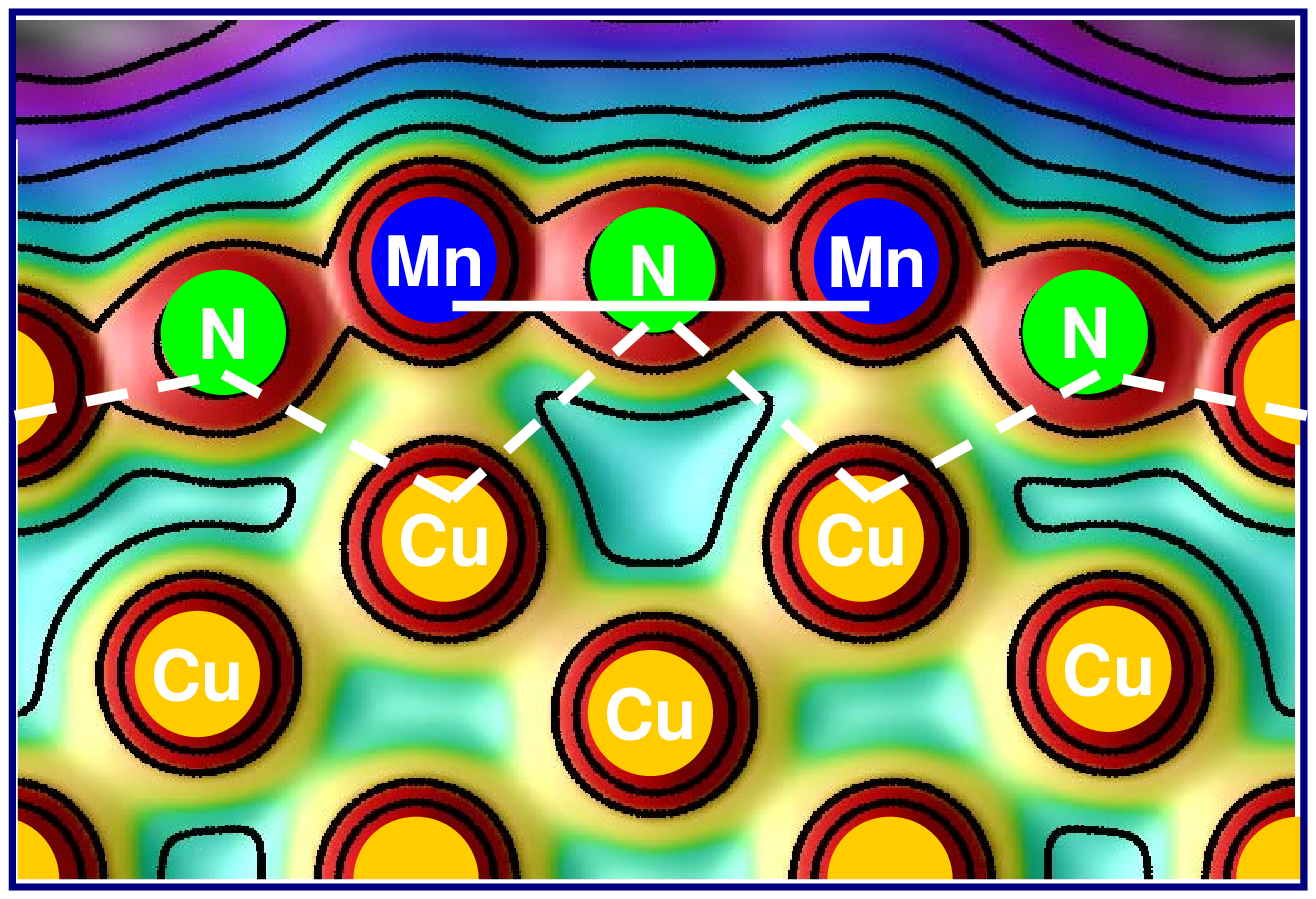}
\end{center}
\caption{\label{fig-Mn2CuN-Cusite-rho} Electron density contour of a Mn
dimer at the Cu site of the CuN surface along the N-Mn-N raw and the out-of-plane direction.
The white solid line shows our proposed coupling path between Mn spins.
The dashed shows how corrugated the clean CuN surface becomes in the presence of Mn adatoms.}
\end{figure}

The electron density contour of the N-site Mn dimer in
Fig.~\ref{fig-Mn2CuN-Nsite-rho} shows a structure completely different
from Mn on the Cu site in Fig.~\ref{fig-Mn2CuN-Cusite-rho}. The Mn dimer on the Cu site forms a chain-like
structure bridged by the significantly lifted middle N atom, while
on the N site the Mn is attached to the surface like a crown. The binding
structures of the Mn atoms strongly suggest that the Mn spins are
coupled through the N atoms. The electron density contours
indicate that the Mn dimer at the Cu site has a coupling path
considerably shorter than when in the surprisingly different
structure at the N site. We propose that this explains why the
exchange coupling $J$ measured by STM for the Cu-site Mn dimer has
a value twice that of the N-site.


\begin{figure}
\begin{center}
\includegraphics[keepaspectratio,width=6.5cm]{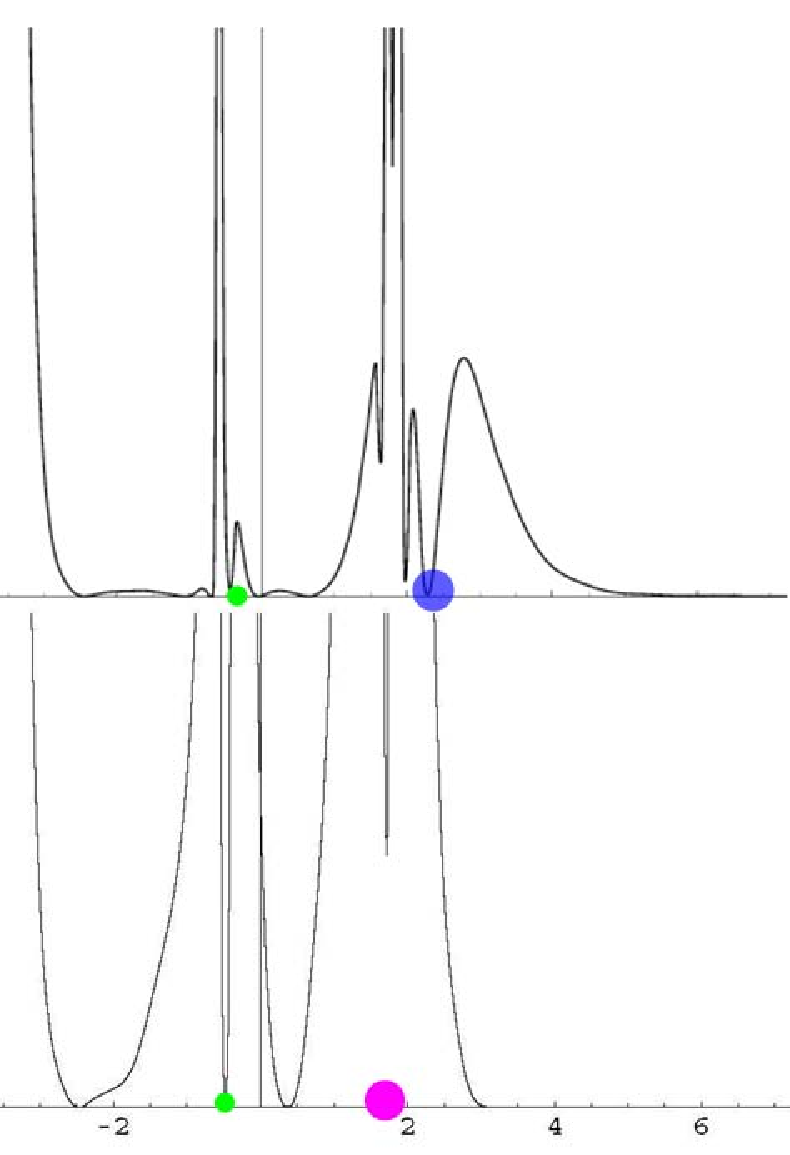}
\end{center}
\caption{\label{fig-MnCo-wfn} LDOS($\epsilon_{F}$) along the
out-of-surface direction through the
adatoms Mn (the larger, solid blue circle in the upper plot)
and Co (the larger, solid purple circle in the lower)
on the CuN surfaces. The smaller, solid green circles are the Cu atoms underneath the
adatoms. The origin is chosen at location of the surface Cu atom of
the clean CuN surface, and the vacuum corresponds to positive values of $z$.}
\end{figure}

\begin{figure}
\begin{center}
\includegraphics[keepaspectratio,trim=5cm 2cm 1cm 4cm,clip,width=8cm]{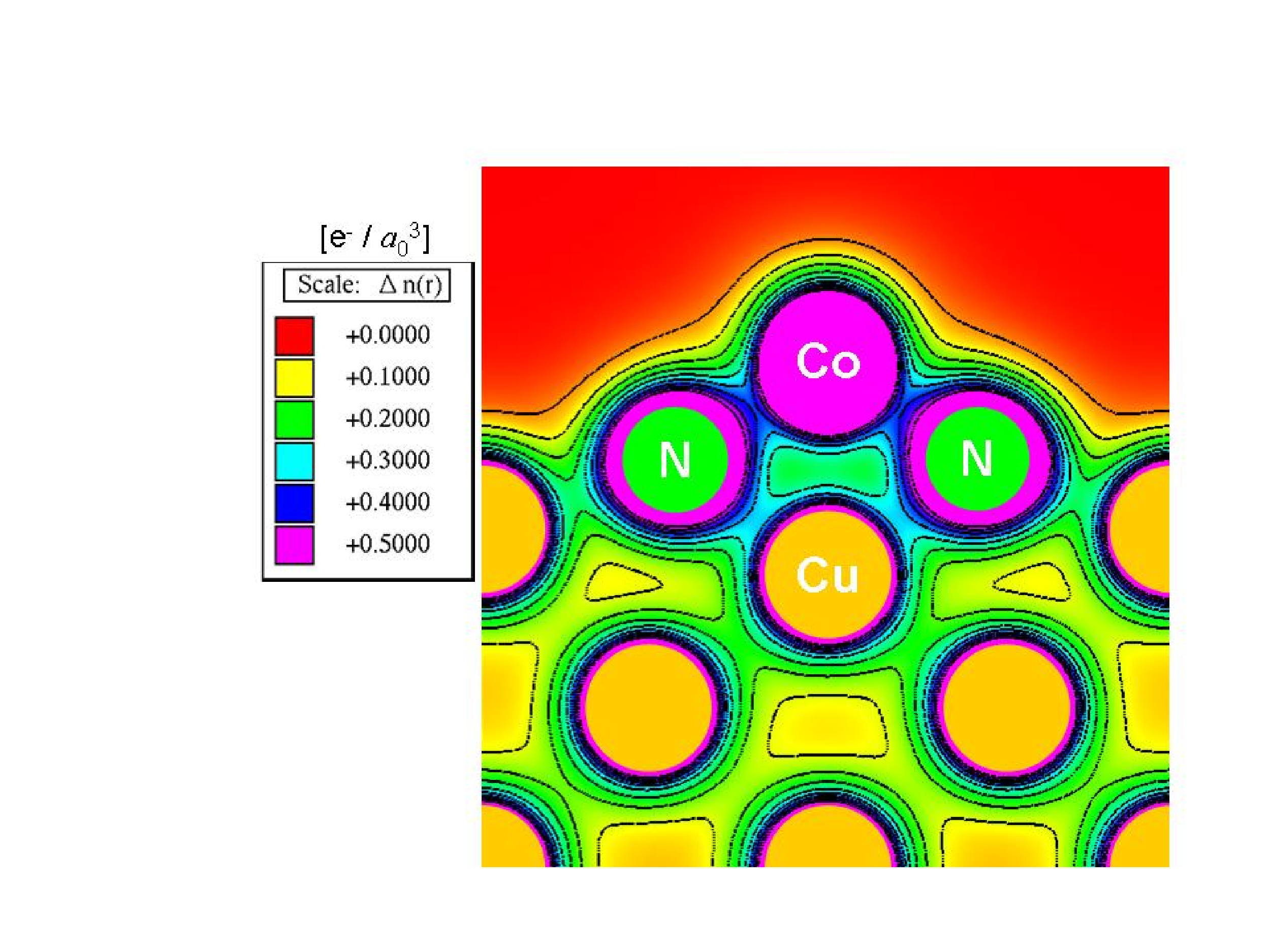}
\end{center}
\caption{\label{fig-CoCuN-rho} Electron density contour of a
single Co on the CuN surface along the N-Co-N raw and the
out-of-plane direction.}
\end{figure}

Co atoms on the CuN surface behave quite differently from Mn as
experiments \cite{CoNaturePhysics,CoPRL} show. Co displays a Kondo effect,
while Mn does not.
The relaxed structure via our calculation shows Co settles lower
in the surface than Mn (see Fig.~\ref{fig-MnCo-wfn}),
and so interacts more with the conduction electrons.
We also compare the surface LDOS with Co and Mn
as in Fig.~\ref{fig-MnCo-wfn}, and find that there is more LDOS between
Co and the Cu underneath it than for Mn. This fact can also be seen by comparing
the charge contour plots of these two systems (see Fig.\ref{fig-MnCuN-rho} and \ref{fig-CoCuN-rho}).
Such substantial LDOS near Co provides
the conduction electrons needed for a Kondo effect to happen.

\begin{figure}
\begin{center}
\includegraphics[keepaspectratio,width=9cm]{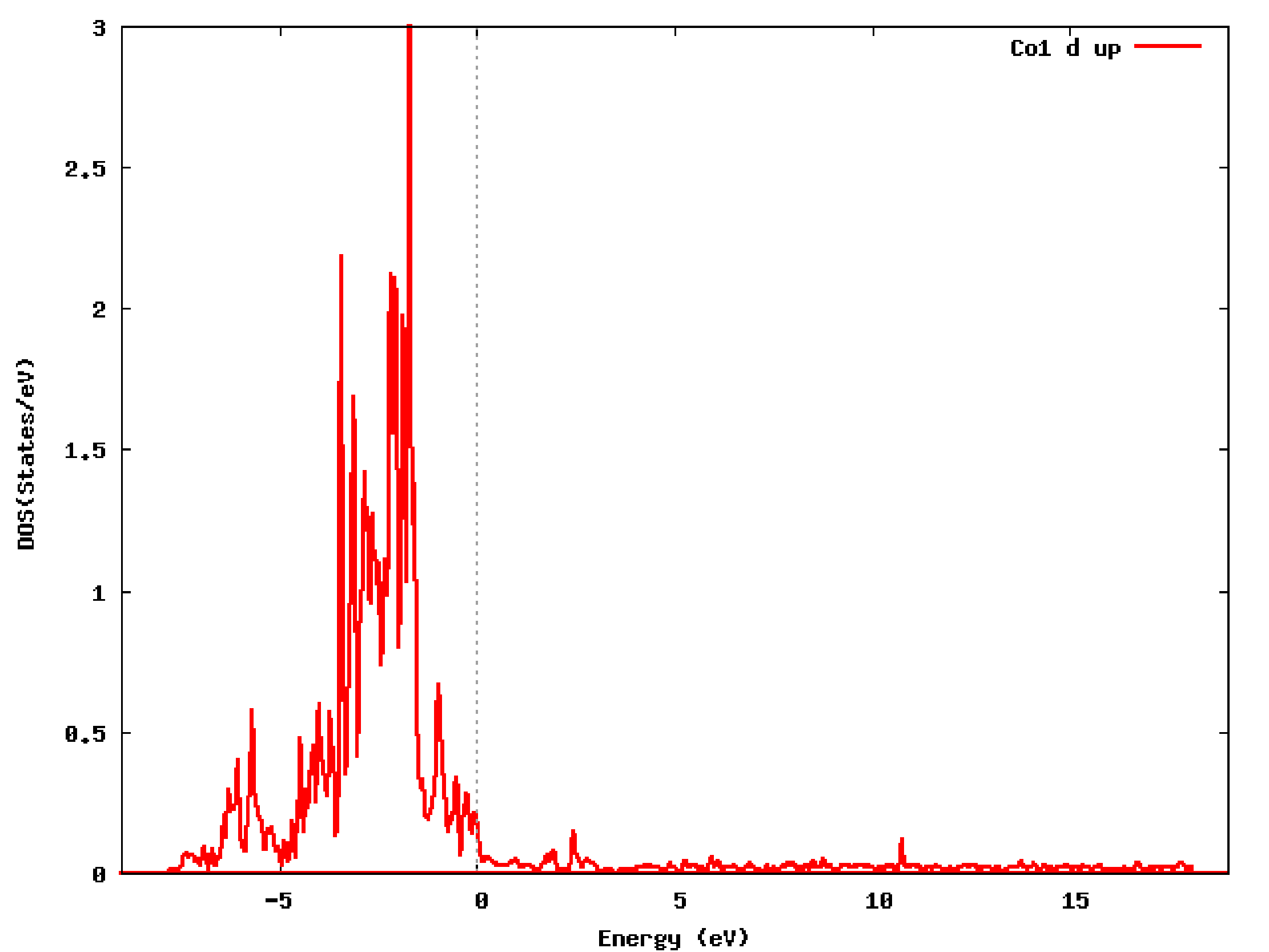}
\end{center}
\caption{\label{fig-Co-spin-up} The Co $3d$ spin-up total density of states on the CuN surface.}
\end{figure}

\begin{figure}
\begin{center}
\includegraphics[keepaspectratio,width=9cm]{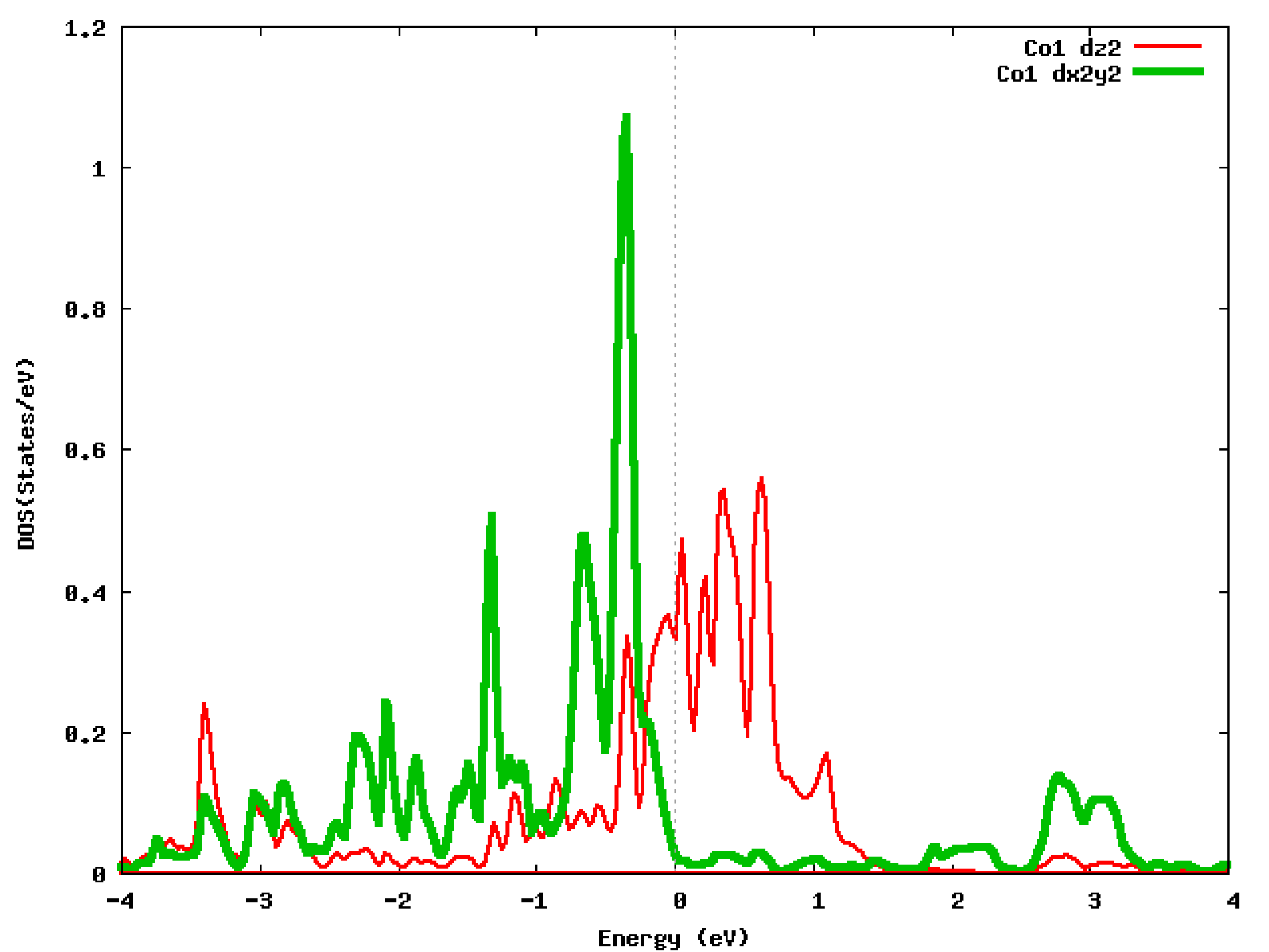}
\end{center}
\caption{\label{fig-Co-spin-down-z2} The Co $3d$ spin-down densities of states of
the $d_{z^2}$ and $d_{x^2-y^2}$ subshells on the CuN surface.}
\end{figure}

\begin{figure}
\begin{center}
\includegraphics[keepaspectratio,width=9cm]{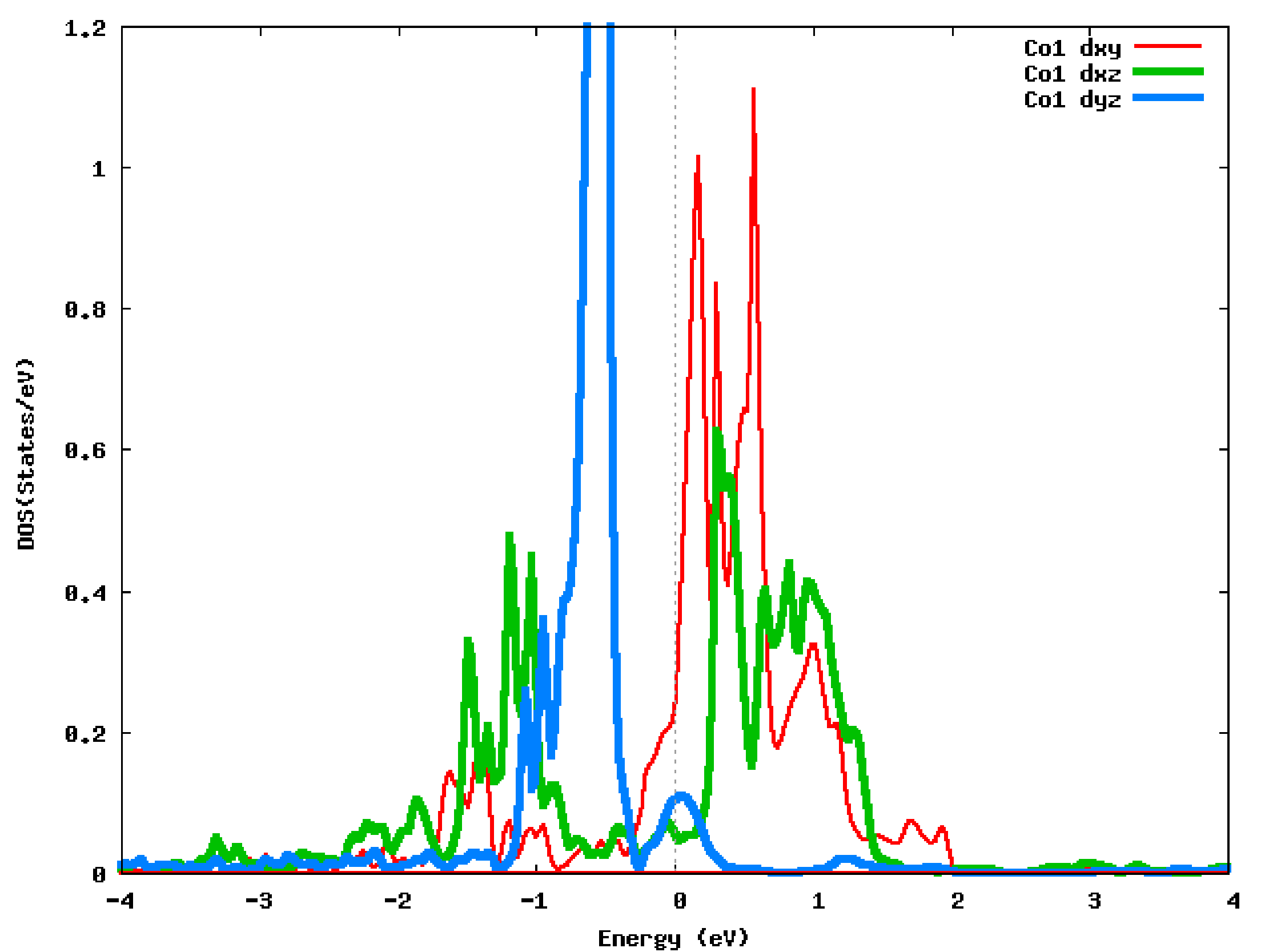}
\end{center}
\caption{\label{fig-Co-spin-down-xy} The Co $3d$ spin-down densities of states of
the $d_{xy}$, $d_{xz}$, and $d_{yz}$ subshells on the CuN surface.}
\end{figure}

To find the Co spin from our calculation, we plot the densities of states of
the $3d$ Co on the CuN surface as in Fig.~\ref{fig-Co-spin-up},
\ref{fig-Co-spin-down-z2}, and \ref{fig-Co-spin-down-xy}.
One clearly sees that the spin-up total density of states and
the spin-down $3d_{z^2}$ and $3d_{xz}$ ones are all occupied,
while the rest are majority unoccupied. This density-of-state analysis gives
$S=3/2$ for Co on the CuN surface by approximating the Co $3d$ in terms of
an atomic-like electron configuration of 5 spin-up and 2 spin-down electrons.
Another interesting point is to compare the $U$ values of Co on Au(111) and
this CuN/Cu(100) surface since Co/Au(111) \cite{Kondoscience} is one of
the most extensively studied surface Kondo systems.
The on-site Coulomb repulsion of Co on Au(111) was extracted to be $2.8$ eV from
a previous first-principles calculation \cite{Ujsaghy}.
The present study has obtained $U=0.8$ eV for Co on the CuN surface.
The substantial difference of Co $U$ of the two systems can be explained
in the way that Co surrounded by N is more positively charged than that on Cu(111),
so adding an electron into Co on CuN is easier because a Co ion attracts an electron
more strongly.

\section{Conclusions}

In summary, we have calculated the electronic structures of
novelly engineered spin systems. The precise atomic charges and
positions of those systems, not accessible by experimental
techniques, are determined by structure relaxation and Bader
analysis respectively in our calculations. The charge analysis
shows that the Mn-N bond formed by Mn adsorbed on the CuN surface
has stronger bond polarity than the Cu-N bond. The presence of Mn
gives rise to substantial rearrangement of the atomic structure:
the Mn atoms at the Cu sites perturb their surrounding atomic positions,
while those at the N sites do not. The calculated $J$'s agree
excellently with the STM measurements for two different Mn binding sites.
Such agreement serves as a touchstone of DFT's
future predictability in similar systems, and is important in
searching for a desired $J$ (e.g. large value or ferromagnetic) for
device applications, with the goal of avoiding multiple
experimental trials. The electronic structures of the Co atoms on
the same surface is also calculated. From that we explain why Co has
Kondo effect while Mn doesn't. We also find the Co spin to be $S=3/2$,
in agreement with the STM's indirect derivation \cite{CoNaturePhysics}.
The on-site Coulomb is calculated to be $U=0.8$ eV, much smaller
than that of the popular surface Kondo system Co/Au(111), which we
explain by the polarities of Co to its nearest neighbor atoms.

We thank C.~F.~Hirjibehedin, C.~P.~Lutz, and A.~J.~Heinrich for
stimulating discussions, and the technical help of the IBM Almaden Blue Gene
support team. C.~Y.~Lin acknowledges supports from the Taiwan National Science Foundation,
the Taiwan National Center for High-performance Computing, and the Taiwan National
Center for Theoretical Sciences (South).

\end{document}